\documentclass[twocolumn,showpacs,aps]{revtex4}

\usepackage{graphicx}

\draft

\def\lapproxeq{\lower .7ex\hbox{$\;\stackrel{\textstyle <}{\sim}\;$}}

\begin{document}

\title{Propagation of ultra-high energy protons in regular extragalactic
       magnetic fields}
\author {Todor~Stanev$^1$, David~Seckel$^1$ \& Ralph~Engel$^2$}
\address {$^1$
Bartol Research Institute, 
University of Delaware, Newark, DE 19716, USA
}
\address {$^2$ Forschungszentrum Karlsruhe, Institut f\"{u}r Kernphysik,
  Postfach 3640, 76021 Karlsruhe, Germany}

\widetext
{
\vspace*{5truemm}
\begin{abstract}
We study the proton flux expected from sources of ultra 
high energy cosmic rays (UHECR) in the presence of regular extragalactic 
magnetic fields. It is assumed that a local source of ultra-high energy 
protons and the magnetic field are all in a wall of matter 
concentration with dimensions characteristic of the supergalactic plane. 
For a single source, the observed proton flux and   the local 
cosmic ray energy spectrum depend strongly on the strength of the field,
 the position of the observer, 
and the orientation of the field relative to the observer's line of 
sight.
 Regular fields 
also affect protons emitted by sources outside the local magnetic
fields structure.
 We discuss the possibility that such effects could contribute to
 an explanation of the excess of UHECR above $5.10^{19}$ eV,
 and the possibility that sources of such particles may be missed
 if such magnetic fields are not taken into account. 
\end{abstract}
 
\pacs{98.70.Sa, 13.85.Tp, 98.80.Es, 98.65.Dx}
}
\date{\today}

\maketitle

\narrowtext

\section{Introduction}
  
The current observations of ultra-high energy cosmic 
rays~\cite{NagWat00} do not allow firm conclusions on the 
existence of the
cut-off of the cosmic ray spectrum between 10$^{19}$ and 10$^{20}$ eV,
as proposed by Greisen and Zatsepin\&Kuzmin~\cite{GZK} (GZK).
 The cut-off should exist if the the source distribution of UHECR
 were isotropic and homogeneous because of photoproduction interactions
 on the microwave background. There also should be~\cite{BerGrig} a
 small pile-up just before the cut-off and a secondary small
 dip in the spectrum due to electromagnetic pair production on
 the same target.
 While some of the highest statistics experiments~\cite{HiRes_n}
 see similar features in their data,
 others~\cite{AGASA} observe a spectrum that extends well beyond
 10$^{20}$ eV.

 The possible absence of the GZK cut-off in the observed UHECR
 suggests that local sources contribute a significant fraction
 of the observed UHECR. The spectrum that could be derived
 by the world UHECR statistics can indeed~\cite{BBOlinto}
 be fit fairly well with a combination of homogeneous and local
 source distributions. The problem with this
 solution is that the cosmic rays above 4.10$^{19}$ eV, which
 should not scatter much in weak random magnetic fields, do not display
 strong large scale anisotropy. The observed small scale
 clusters~\cite{asclust} do not point at any known nearby astrophysical
 system, or at regions of  increased matter density. 

 The isotropy of UHECR can in principle be explained by
 scattering in large scale magnetic fields. For example,
 one suggestion is~\cite{Ahnetal} that there is a sizable Galactic halo,
 similar to the heliospheric one, that isotropizes UHECR protons.

 Understanding the influence of the large-scale magnetic fields on
 the cosmic ray propagation is vital for obtaining reliable information
 on the sources. The cosmic ray injection power of the sources,
 needed to maintain the observed highest energy cosmic ray flux, depends
 directly on the large-scale magnetic fields in the vicinity of the Galaxy.
 Magnetic fields can not only change the locally observed intensity but
 also the energy spectrum of UHECR~\cite{Tanco98,SEMPR,Deligny}. 
Depending on the
 typical field strength, protons with energies greater than $10^{20}$
 eV are hardly deflected during propagation whereas, for example,
 particles of $E=10^{18}$ eV
  may have a diffusive propagation pattern. In such models all cosmic 
 rays could be local, with observed spectra strongly influenced by field 
 geometry and source distributions.

 Several earlier papers~\cite{Lemoineetal97,Tanco98,Sigletal99,Ideetal01}  
 have studied the acceleration and propagation of UHECR protons in 
 intense magnetic structures. 
 Most of these papers attempt to create quasi-realistic scenarios,  
making the straight-forward understanding  of the involved
 physical processes difficult.  We take the opposite approach and
 introduce relatively simple, yet qualitatively different,
 magnetic field geometries and deal
 with a single cosmic ray source. This makes it possible to
directly follow and understand the consequences of regular  large scale
 magnetic fields.  We first assume that both the nearby UHECR source 
 and the Galaxy are inside a wall with high concentration of
 matter (supergalactic plane (SGP)), which also creates a large
 scale magnetic field structure.
 Restricting our considerations  to a nearby source at 20 or 40 Mpc,
 we study the influence of 
 different  magnetic field configurations on the particle densities,
 arrival directions and energy spectra.
 
 Another scenario that we investigate is that of a source well outside
 the local magnetic field environment. In such a case the effects on the
 `detected' UHECR are different, but can also be understood on the basis
 of the same processes that affect the local source scenario.

 Before proceeding, we emphasize the following points to be kept in mind 
 as our results are presented. The effects of our nominal field models 
 qualitatively divide into two energy regions which, coincidentally or 
 not, correspond to UHECR with energies above and below the GZK cutoff 
 ($10^{19.5}$ eV). In the high energy regime the primary effect is to 
 modestly change the direction of UHECR. The relevant experimental 
 question is, ``Can the isotropy of super-GZK UHECR be affected?'' At lower 
 energies particle propagation becomes diffusive, so that the relative 
 geometry of source and observer in the regular magnetic field can 
 strongly influence the UHECR spectrum and intensity. Here, the relevant 
 question becomes, how do magnetic fields affect ones ability to infer 
 source properties by measurements of the spectra? Third, since
 the transition energy is near the GZK cuto-off, the apparent strength 
 of the GZK spectral feature may be affected by the presence of large
 scale non-random magnetic fields.

 The outline of the paper is as follows: We discuss the supergalactic 
 plane and field geometry in Section 2, which also gives some details of 
 the calculation. The results on the particle densities within the
 20 Mpc sphere as a function of the magnetic field geometry are
 presented in Section 3. The energy spectra and directions
 of particles leaving the 20 Mpc sphere at different angles from the
 magnetic field direction are given in Section 4. In Section 5
 we discuss the boundary conditions of the simulation - the distance from
 the source to the observer, the possible existence of an external source
 and the time dependence of the UHE proton spectra in the case of an
 impulse injection of UHECR. The  paper concludes with a discussion
 in Section 6.  

\section{Method of calculation}

\begin{figure}[t] 
\centerline{\includegraphics[width=5.5cm]{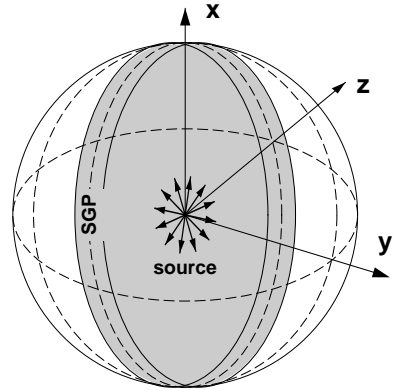}}
%
\caption{ Geometry of the calculation in Cartesian coordinates. The
 supergalactic plane is infinite in $x$ and $z$ directions. Protons are
 injected at the origin and recorded when they cross the 20 Mpc sphere.
\label{draw}
}
\end{figure}

 We consider a single source in the supergalactic plane at
 a distance of 20 Mpc, compatible with the distance to the local 
 Supercluster. This setup is motivated by the assumed
 matter concentration within the plane and the small number
 of powerful astrophysical systems in our cosmological neighborhood. 
 The geometric structure of the calculation is illustrated
 in Fig.~\ref{draw}.

\subsection{Magnetic field geometry} 

 It is convenient to define the supergalactic plane as an
 infinite plane coinciding with the $y=0$ plane in Cartesian coordinates. 
 The magnetic field strength of the large scale field is assumed to be 
 constant ($B_0 = 10$ nG) up to a distance of 1.5 Mpc on both sides
 of the plane, i.e. the SGP has a width of 3 Mpc.
 At larger distances the regular magnetic field decays exponentially
 with a decay length of 3 Mpc.

 We assume that in addition to the regular magnetic field there
 is a turbulent field. We take the strength of this random field to
 be one half that of the regular field, but never less than
 1 nG. The random field is represented by a Kolmogorov expansion
 on three scalelengths of 1, 0.5 and 0.25 Mpc. For a discussion
 of the implementation of the random field see Appendix B of
 Ref.~\cite{SEMPR}. 

 In general it is assumed that the direction of the regular field
 coincides with the gravitational flow of matter, i.e. towards the
 supergalactic plane, and possibly along the SGP towards a higher
 concentration of matter in nodes, i.e.  clusters of galaxies.
 This general idea is supported by some of the simulations of large
 scale structure formation~\cite{Ryuetal98}
 We use three possible implementations of this idea:\\
 {\bf SGP\_A}: The large scale field is parallel to the SGP ($B_z=B_0$,
 $B_x=B_y=0$).\\
 {\bf SGP\_B}: The large scale field is orthogonal and points towards
 the SGP ($B_y= B_0$ for negative $y$ and
 $B_{y}=-B_0$ for positive $y$, $B_x=B_z=0$).\\
 {\bf SGP\_C}: The field is orthogonal to SGP at distances greater 
 than 1.5 Mpc from it (see SGP\_B), and 
 parallel to it at closer distances (see SGP\_A).

\begin{figure}[t] 
\centerline{\includegraphics[width=7.5cm]{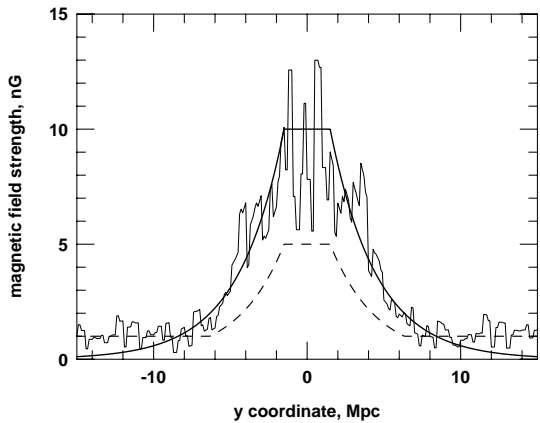}}
%
\caption{Realization of $B_\parallel$, the magnetic field component 
parallel to the large scale regular field $B_0$. Depending on the model 
$B_\parallel$ may correspond to $B_z$ (SGP\_A), $B_y$ (SGP\_B) or switch 
between the two at 1.5 Mpc (SGP\_C). The heavy and dashed curves show 
the magnitude
of the regular and turbulent magnetic field components, respectively.
\label{f:magfield}
}
\end{figure}

 A realization of our magnetic field model is illustrated in 
 Fig.~\ref{f:magfield}. Within the SGP and out to a distance
 $y_{reg}\sim 8$ Mpc the regular field dominates, but outside
 this distance the field is essentially random. Between 1.5
 and 8 Mpc there is a systematic gradient to the magnetic field
 strength. 

 The implementation of the regular field as an infinite plane
 does not allow us to close the magnetic field lines to satisfy
 $\vec\nabla \cdot \vec B = 0$. This does not affect our results,
 because we are propagating the UHECR protons to distances 
 of 20 Mpc, assuming that the field lines are closed on a
 larger scale.

\subsection{Source spectrum and propagation}

 We restrict our considerations to protons as UHECR. Similar
 effects due to magnetic fields are obtained for heavier nuclei,
 if the energy is rescaled correspondingly and the differences
 in the energy loss are accounted for. The particles are
 injected at the origin of the coordinate system within a cone
 centered about the positive $z$ direction. The injection cone
 is described by its opening angle $\theta$ which may be varied
 according to the intended purpose. We prefer to inject 
 particles into a cone of solid angle 2$\pi$ because in this case
 we can easily identify the particles that propagate in directions
 opposite to their injection direction.

 Protons are injected with an energy spectrum
\begin{equation}
N(E) = \left(\frac{E}{E_0}\right)^{-1.25}
\exp\left\{- \frac{E}{E_{\rm c}}\right\}\; ,
\label{espectrum}
\end{equation}
 with $E_{\rm c}$ = 10$^{21.5}$ eV and then usually weighted
 to represent an $E_0^{-2}$ spectrum with the same exponential
 cut-off. We will warn the reader for images that present results
 with the unweighted injection spectrum.

 The particle trajectories are integrated numerically with 
 a stepsize of 10 kpc. First the probability  for photoproduction
 interaction is calculated,  and if such an interaction takes
 place  the photoproduction interaction is simulated by the
 event  generator SOPHIA~\cite{SOPHIA}.  The energy loss due to
 $e^+e^-$ pair production is calculated  at each step and treated
 as a continuous process. The particle direction is calculated
 at each step as a function of the local magnetic field strength
 and direction. Neutron production and decay is taken into account. 
 Each injected proton is followed until it  crosses a 20 Mpc 
 sphere around the injection point, or a propagation time of
 $1.3 \times 10^9$ yrs  has elapsed.
  The adiabatic losses are accounted for at every step,
  but starting the propagation at redshift of 0.005, which corresponds
  to a distance of 20 Mpc for light propagation.We expect the errors
  due to this procedure to be less than 10\%, significantly smaller
  than the uncertainties in the magnetic field strength and configuration. 
 Different numbers of particles
 are injected in different runs,  depending on the size of the
 solid angle at injection and the collection area. For injection
 in one hemisphere ($\theta=90^\circ$)  the 
 typical number of calculated trajectories is $10^{6}$.

 Although we solve explicitly for particle propagation in the magnetic 
 fields, we qualitatively expect the following behaviours:\\
 a) At the highest energies protons propagate in nearly straight lines. The 
 gyroradius is given in Mpc by $r_L \simeq E_{18}/B_{-9}$
 where $E_{18}$ is the proton energy in EeV, and $B_{-9}$ is the magnetic
 field in nG. Thus, for a
 10 nG field, particles with energies above $10^{19.5}$ eV will pass 
 through the SGP with only a modest deflection. Above $10^{20}$ eV 
 particles propagating as neutrons will increase the effective diffusion 
 length for UHECR.\\
 b) Lower energy particles will be strongly affected by magnetic fields. 
 Outside $y_{reg}$ the fields are essentially random, and propagation is 
 isotropic and diffusive on a scale comparable to $r_L$. \\
 c) For $y < y_{reg}$ the diffusion tensor is not isotropic. Diffusion is 
 much easier along field lines than perpendicular to them, but the scaling 
 is complicated in the region where $r_L$ is comparable to the coherence
length $l_{coh}$ of the turbulent component of the magnetic
 field~\cite{ZankMBM98}. 
 The random component of the magnetic field yields a coherence length of 0.39 
 Mpc\cite{SEMPR}. We estimate that $\lambda_\parallel$, the diffusion 
 length parallel to $B_0$, increases from $< 10$ Mpc to $> 100$ Mpc as the 
 proton energy increases from $10^{18.5}$ to $10^{21}$ eV. 
 Correspondingly, diffusion across field lines is quite inefficient, 
 $\lambda_\perp \ll 1$~Mpc. \\
 d) Although diffusion across field lines is slow, due to the large 
 gradients in our field model there may be significant $B \times \nabla B$ 
 drifts.

\section{UHECR density in different magnetic field configurations}
 
\begin{figure}[t] 
\centerline{\includegraphics[width=8.5cm]{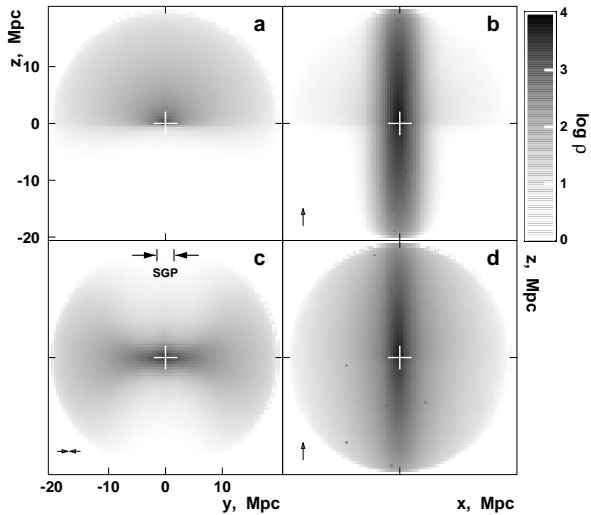}}
\caption{Particle density projections. a) projection on the \protect$yz$
plane for random magnetic field of average strength 1 nG; 
b) projection on the \protect$yz$ plane for model SGP\_A; c) projection
on the \protect$yz$plane for model SGP\_B; d) projection on the \protect$xz$
plane for model SGP\_A. The symbol at the lower left corner of each
panel indicates the direction of the regular field.  
\label{fig1}
}
\end{figure}

 Fig.~\ref{fig1} shows the projection of the particle density
 (i.e. density of trajectories) on the $yz$ (panel a,b,c) and $xz$ planes
 (panel d) for different magnetic field models. Protons
 are injected with the spectrum~(\ref{espectrum}) in 2$\pi$ steradian
 with $\cos \theta$ between 1 and 0 in direction of positive $z$.
 The minimum injection energy is $10^{18.5}$ eV.
 At each propagation step of 10 kpc the position of the particle is
 projected to the respective plane. The particle density is dominated
 by the lowest energy particles not only because of the steep
 injection spectrum, but also because these particles are scattered
 more by the magnetic field and have larger pathlengths to reach the
 20 Mpc sphere.

  Fig.~\ref{fig1}a shows the density projection on the $yz$ plane in the
 absence of any regular field. Only the random 1 nG field is present. Although
 the injection is restricted to the $+z$-hemisphere some of the 
 protons scatter
 backwards and partially populate the backward hemisphere. 

Panel b) shows the projection on the $yz$ plane for the model SGP\_A, 
where the magnetic field is parallel to the supergalactic plane. One 
sees the enhancement of the proton density around the SGP. These are 
also mostly low energy protons trapped by the magnetic field. The 
`halo' that fills the forward hemisphere consists of high energy 
particles  that do not deviate significantly from the direction of 
injection.

Fig.~\ref{fig1}c is for model SGP\_B where the magnetic field direction  
follows the flow of matter to the SGP.  The pattern is almost exactly the 
opposite to that in panel  b) -- at small distance from the injection 
point the protons are constrained to follow $B_0$ and in this case soon 
leave the SGP. When they reach $y \sim y_{reg}$ the diffusion 
becomes isotropic. There is also a small concentration along the SGP 
close to the injection point from higher energy particles injected near 
$\theta =0$. The density of  particles crossing the 20 
Mpc sphere inside the supergalactic plane is  very small and consists 
mainly of the highest energy protons.

Fig.~\ref{fig1}d shows the projection on the $xz$ plane for SGP\_A.  
This is the plane that contains the magnetic field. From the perspective 
of cross-field diffusion the extended distribution in the $x$-direction 
is somewhat puzzling. Upon investigation, a correlation between $x$ and 
$y$ positions indicates that cross-field movement in the $x$ direction is 
in fact particle drift due to the gradient of the magnetic field.

\begin{figure}[t] 
\centerline{\includegraphics[width=8.5cm]{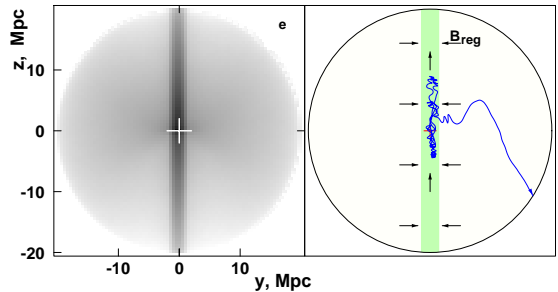}}
\caption{ Right-hand panel: projection of a proton trajectory on the
 $yz$ plane. The arrows show the magnetic field direction outside
 and inside the SGP. Left-hand panel:  Particle density projection
 on \protect$yz$ plane for magnetic field model SGP\_C. 
\label{fig2}
}
\end{figure}

 Fig.~\ref{fig2} shows the projection of the particle density on the 
$yz$  plane for model SGP\_C and the trajectory of a typical low energy  
particle ($E<10^{19}$ eV) projected on the same plane. The particle  
is injected at the origin and is immediately trapped in the SGP  field. 
It diffuses back and forth along the plane with $\lambda_\parallel$ of 
order 5-10 Mpc (by eye), until it eventually heads off in the $y$ 
direction, presumably influenced by a strong random field. Once it leaves 
the  SGP the field direction in the SGP\_C model switches from $z$ to 
$-y$  direction. The proton is now trapped in this field, gyrates in it  
for a few Mpc and escapes out of the 20 Mpc sphere.

 The left-hand panel shows a density pattern which qualitatively
corresponds to trajectories such as that in the right-hand panel. 
There is a strong enhancement of
 the particle density in a narrow region coinciding with the SGP itself.
 The enhancement is stronger in the vicinity of the injection point.


 All three models of the magnetic field configuration show
 qualitatively the same propagation behaviour: UHECR of energy
 above about 10$^{19.5}$ eV propagate almost rectilinearly, while
 the lower energy particles concentrate along the magnetic field lines.
 Because of that in the rest of this paper we concentrate on model 
 SGP\_A assuming that, although results would be different in detail,
 the main conclusions will still be robust.

\section{Energy spectra of UHECR leaving the 20 Mpc sphere at
 different locations}

 It is reasonable to expect that the changing particle density as a 
 function  of the magnetic field model, and the location of the observer, 
 would lead  to changes in the observed energy spectrum. For example, from 
 Fig.~\ref{fig1} we see that for model SGP\_A the particle flux below 
 10$^{19}$ eV will mostly escape out the end caps, ($|x|,\,|y| <1.5$ Mpc, 
 $z=\pm 20$ Mpc) an area of order 10 Mpc$^2$. If there were no magnetic 
 field this same flux would escape through the full 2500 Mpc$^2$ surface 
 area of the hemisphere. Thus, an observer in the endcap region sees a 
 flux enhanced by a factor of $\sim 100$, for energies below a few 
 10$^{19}$ eV. Similarly, an observer who sees the source across field 
 lines or lies outside the plane of the galaxy will see a severely reduced 
 flux at low energies. At the same time, particles with energies above 
 10$^{20}$ eV will hardly be bent, and all observers see roughly the same 
 flux independent of their position. Clearly, the observed spectrum of 
 UHECR may be influenced by the presence of large scale magnetic fields. 
 Correspondingly, efforts\cite{Waxman95} to normalize the injection power 
 of UHECR sources by comparing to the observed spectrum at $E \sim 
 10^{19}-10^{19.5}$ eV are fraught with uncertainty.

 To investigate this question in more detail we perform simulation
 runs in which we record the information for all particles that
 leave the 20 Mpc sphere.
\begin{figure}[t] 
\centerline{\includegraphics[width=5.5cm]{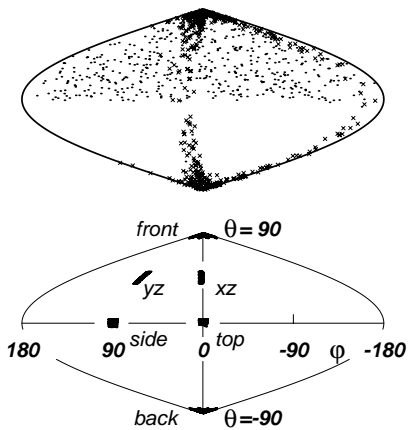}}
\caption{ Upper panel: location of the exit points from the
 20 Mpc sphere for protons of exit energy above (points) and
 below (crosses) 10\protect$^{19.3}$ eV. The first 500 protons in
 each of the energy bins are plotted without weighting. 
 Lower panel - locations of the six panels defined in the text.
 The vertical centerline is at y=0, x$>$0 and the outer boundary is at
 y=0, x$<$0.
\label{caps_intro}
}
\end{figure}
 The upper panel of Fig.~\ref{caps_intro} shows the exit points
 of protons subdivided in two energy groups - above and below
 10$^{19.3}$ eV. The exit points reflect the particle densities
 shown in Fig.~\ref{fig2}. 
The exit points for high energy particles map out the $2\pi$ beam of 
injected particles. The lower energy particles include two populations. 
The main population consists of particles that stay within the SGP, 
diffuse along field lines of the large scale magnetic field, and exit in 
either the $+\hat{z}$ or $-\hat{z}$ direction. The second population is 
particles that propagate to the edge of the SGP. These particles 
experience not just a large scale $\vec{B}$, but also a large scale 
$\nabla {B}$, and so drift across field lines. The orientation is such that 
protons on the positive $y$ side of the SGP drift in the $+x$ direction, 
whereas particles on the negative $y$ side of the SGP drift in the $-x$ 
direction.
When projected onto the 20 Mpc sphere, the exit points for 
these particles define two bands, one of which intercepts the equator at 
$\phi \simeq 10-20$ degrees and the other at $\phi \simeq 190-200$ degrees.
 Note that due to the drift process, the exit points
 tend to be on the maximum $|y|$ excursion of the particle orbit.

The top panel of Fig.~\ref{caps_intro} shows a global picture of the exit 
points for two energy bins, but it does not give an adequate picture of 
the diversity of spectra that may be seen by an individual observer. 
Accordingly, we define six ``caps" on the surface of the 20 Mpc sphere, 
each of area 3$\times$3 Mpc$^2$, at the following locations:\\
 \hspace*{7truemm}{\em front} at positive $z$, $|x|,|y|$ less than 1.5 Mpc\\
 \hspace*{7truemm}{\em back} at negative $z$, $|x|,|y|$ less than 1.5 Mpc\\
 \hspace*{7truemm}{\em side} at positive $y$, $|x|,|z|$ less than 1.5 Mpc \\
 \hspace*{7truemm}{\em top} at positive $x$, $|y|,|z|$ less than 1.5 Mpc \\
 \hspace*{7truemm}{\em yz} in the forward hemisphere centered at 
 (0, 20/$\sqrt{2}$, 20/$\sqrt{2}$)\\
 \hspace*{7truemm}{\em xz} in the forward hemisphere centered at 
 (20/$\sqrt{2}$, 0, 20/$\sqrt{2}$)\\
The terminology ({\em front, back}, etc.) reflects the orientation 
illustrated in Figure \ref{draw}.

\begin{figure}[t] 
\centerline{\includegraphics[width=8.5cm]{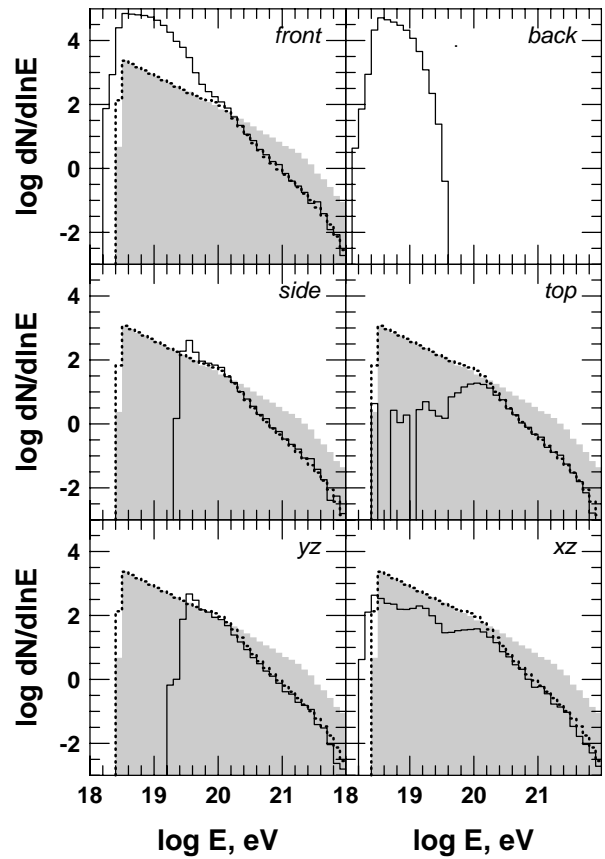}}
\caption{ Energy spectra of protons that cross the 20 Mpc sphere within
 9 Mpc\protect$^2$ caps. The shaded histogram is the injection
 (\protect$E^{-2}$) spectrum and the dotted histogram shows spectra
 propagated at 20 Mpc in the absence of magnetic fields.
 There is no injection spectrum within
 the solid angle of the {\em back} cap. See the text for the locations
 of the six caps. 
\label{fig3}
}
\end{figure}

 Fig.~\ref{fig3} shows the spectra of the particles that leave
 the 20 Mpc sphere through these six caps. The shaded area shows the 
 injected particle spectrum for the same solid angle.
 The dotted histogram shows the result of propagation of protons
 in the absence of magnetic field but including energy loss. 
 The spectrum of particles leaving the sphere through the {\em front}
 cap shows the expected enhancement at energies below 10$^{20}$ eV.
 At 10$^{19}$ eV the flux is enhanced by a factor of about 60.
 The decrease of particles above 10$^{20.2}$ eV is due to the 
 energy loss during propagation, as can be seen by comparing
 to the dotted histogram. Although 20 Mpc is a cosmologically
 negligible distance, it is at least a factor of three higher than
 the mean free path for photoproduction $\lambda_{phot}$ at energies
 above 10$^{20}$ eV, which reaches a minimum of about 3.8 Mpc at
 10$^{20.5}$ eV~\cite{SEMPR}.

 The {\em back} cap only sees 
 backscattered protons of energy below 10$^{19.6}$. In reality, for
 4$\pi$ injection both the {\em front} and the {\em back} caps
 would see the sum of the shown two distributions. This would almost
 double the excess of low energy particles seen by observers connected to 
 the source by lines of the large scale magnetic field.

 All other caps show a deficit of lower energy particles. The particle
 spectra cut off at about 10$^{19.3}$ for the {\em side} and {\em yz}
 caps. Interestingly, there is a narrow enhancement in the spectrum at
 10$^{19.5}$ eV. At this energy the {\em side} cap accepts protons
 that gyrate by either $90 - \delta$ or $90 + \delta$ degrees, and 
 the exposure of the cap is effectively doubled. 
 The two caps in the $xz$ plane above the particle source show a much
 more complicated spectra.
 The break in the spectrum occurs at higher energy than for the
 {\em side} and {\em yz} caps due to a longer path within the high
 field region of the SGP. At lower energies the spectrum is filled 
 by a combination of particles with modest bending outside the SGP 
 and particles that drift along the edge of the SGP.
 Note that the injection spectra for the {\em side} 
 and {\em top} caps are lower by a factor of 2 since only half of the
 patch is exposed to the beam.

 Fig.~\ref{fig3} demonstrates how strongly the spectrum of the observed
 particles from a single source depends on the relative directions
 of the observer and the magnetic field lines. Since the normalization
 point for estimates of the UHECR source luminosity is in the vicinity
 of 10$^{19}$ eV one could easily under or overestimate the
 luminosity by orders of magnitude. If there are several nearby sources
 these effects may be somewhat ameliorated due to averaging over
 different geometries~\cite{Tanco98}.

\begin{figure}[t] 
\centerline{\includegraphics[width=8.5cm]{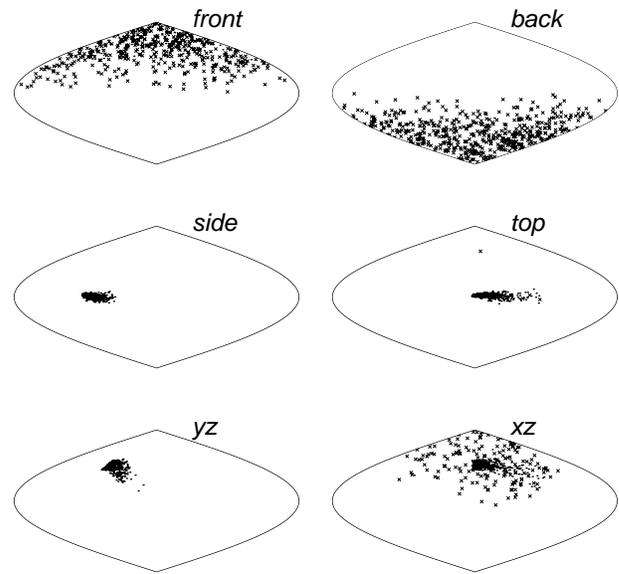}}
\caption{ Scatter plot of the directions of protons leaving the
 20 Mpc sphere through the six caps defined in the text. Protons
 with energy below 10\protect$^{19.3}$ eV are shown with crosses and
 higher energy particles with points. The particles are not weighted
 and misrepresent the ratio of the fluxes in the two groups.
\label{fig4}
}
\end{figure}

 Fig.~\ref{fig4} shows velocity vector maps for the first 500
 protons crossing the six caps. Protons with energy below 10$^{19.3}$
are shown with crosses and above that energy with points.
 The division at 10$^{19.3}$ is suggested by the spectral features
 in Fig.~\ref{fig3}. Note that the entries in these scatter plots
 are not weighted, i.e. they represent events simulated  on the flat
 $E_0^{-1.25}$ injection spectrum. For a more realistic spectrum
 the number of lower energy protons will increase by a factor of 
 about 2.3 and the number of higher energy ones will decrease by a factor
 of about 4 compared to the population of the graph.

  One might expect that, since many particles are contained
 within the supergalactic plane, the magnetic fields will
 `collimate' the bunch and generate an angular anisotropy in
 the {\em front} and the {\em back} caps. Fig.~\ref{fig4} shows
 exactly the opposite picture - the angular distribution of the
 particles leaving through these two caps is very wide, if not
 isotropic. The reason is obvious - these particles that are contained
 by the magnetic field lines gyrate around them and leave the sphere 
 with a variety of velocity vectors. The magnetic field increases
 the particle flux in this direction, but the particles do not point
 at their source. It seems that it would be best to infer the local 
 magnetic field direction from the overall gradient in the particle 
 flux. The situation is similar for 
 both {\em front} and {\em back} caps. The latter one contains only
 three higher energy particles, which a proper weighting would
 eliminate from the statistics of 500. 
 
 All other caps are populated mostly by higher energy protons.
 The {\em side} and {\em yz} caps contain actually no lower
 energy events. These are the positions where the particle
 velocity vectors point best at their source.
 The average deviation from the source direction for these
 caps is about 15$^\circ$ with $\sigma$ of 10$^\circ$.
 These numbers are obtained with the properly weighted distribution
 that decreases significantly the fraction of higher energy events.

 The {\em top} and {\em xz} caps show a more complicated 
 velocity distribution. From the shape of the distribution
 we can conclude that it is very much influenced by drifts
 and bending as the protons approach the cap from a variety of
 angles in the `{\em xy}' plane.
 The protons exiting through the {\em xz} cap consist
 obviously of two separate populations: the higher energy protons with
 relatively small scattering angles and the widely distributed
 lower energy particles that do not retain the memory of their source
 direction.   

\section{The boundary conditions \label{boundary}} 

The discussion so far has focused on a relatively small set of 
simulations, chosen to be simple enough to understand, yet complex enough 
to elicit behaviors which would distinguish models with large scale 
coherent magnetic fields. The general conclusion is that such models show 
effects that would complicate the process of determining source 
properties from observations made at a single point. Having said this, 
the conclusions depend on a fairly small set of models, with particular 
parameter values and choice of boundary conditions. It is interesting to 
know if the conclusions are robust with regard to these choices.

Other than geometry, the relevant parameters are the 3 Mpc thickness of 
the supergalactic plane, the strengths of the regular and random fields, 
and the rate of energy loss for UHE particles. The latter is fixed by our 
knowledge of particle physics and cosmology, but the others are variable. 
Changing parameters will alter the quantitative behavior of the 
simulations, but should not affect the qualitative conclusions. For 
example, we discern a characteristic break in behavior at energies of 
approximately $2 \times 10^{19}$~eV, corresponding approximately to the 
energy at which the gyroradius is equal to the thickness of the 
supergalactic plane. Lowering $B_0$, or decreasing the thickness of the
supergalactic plane, should lower 
the energy above which particles can effectively escape without 
significant bending of their trajectories. The characteristic break would 
remain, but its location would change. 

Similarly, the relative strength of the regular and random fields 
determines the importance of various transport mechanisms for low and 
medium energy particles. As long as the regular field strength exceeds 
the random component, low energy particles will tend to follow the 
regular field within the supergalactic plane. At the edges of the plane, 
transport of low and intermediate energy protons is dominated by $B 
\times \nabla B$ drift in the $+\hat{x}$ direction if $y>0$ and oppositely 
for $y<0$.

Apart from such observations, we will not undertake a study of how 
varying parameters of our model may affect our conclusions, but instead 
focus on the effect of differing boundary conditions. Our 
simulation assumes a constant luminosity point source at the center and 
propagation to the edge of a sphere, after which the particles are 
assumed to escape. Accordingly, we examine three variations, one designed 
to study the boundary condition as particles exit the sphere, one which 
studies a different source geometry and one which studies a different 
source history.

\subsection{Radial size of simulation}

Our assumption that particles escape seems rather simplistic. Surely, a 
realistic model must allow for backscattering, and so an observer on the 
surface of the sphere would see an additional flux of particles we have 
not accounted for. Diffusive backflow will tend to increase the 
density of particles in the simulation and increase their average 
lifetime. On the other hand, particles that leave the simulation due to 
drift are not expected to return. 

 To address these issues, we perform another
 simulation run, where the 20 Mpc sphere is inside a concentric 40 Mpc
 sphere. We record all particles whenever they cross either sphere.
 The proton propagation ends when the particle crosses the 40 Mpc
 sphere, or when its total pathlength is longer than 400 Mpc.
 For the chosen injection spectrum, without weighting, there are on the
 average 1.12 back scatterings per injected proton and 2.08 exits from
 the 20 Mpc sphere. There are obviously some protons that exceed
 the maximum allowed propagation pathlength while inside the 20 Mpc sphere.
 About 90\% of all injected particles leave the 40 Mpc sphere, with
 4\% of the injected protons exceeding the time constraint within the 20 Mpc
 sphere, and another 6\% in the region $20 < r < 40$~Mpc.

\begin{figure}[t] 
\centerline{\includegraphics[width=8.5cm]{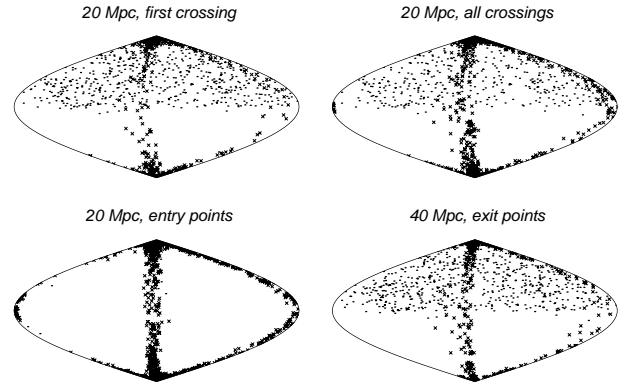}}
\caption{Crossing positions in the run with two enclosed spheres.
 Dots show protons with energy above 10\protect$^{19.3}$ eV, \protect$x$'s
 are for protons below that energy. The particles are not weighted
 and only the first 500 particles in both energy ranges are plotted.
 See text for a better description of the different plots. 
\label{exit_vec}
}
\end{figure}

 Fig.~\ref{exit_vec} shows the proton crossing points for the two spheres. 
 The top lef-hand panel is for protons that exit the 20 Mpc sphere
 for first time. This plot should be identical to the left-hand panel
 in Fig.~\ref{caps_intro} and it is almost identical, within the
 statistical uncertainty for the limited number of plotted points.
 The bottom left-hand panel is for protons that scatter back into
 the 20 Mpc sphere after they have left it. As expected, the diffusion 
population is still concentrated at the two poles of the distribution. 
Note, however, that the offset in $\phi$ for the band of drift particles is 
smaller for the set of reentry intersections than for the exit 
points. This is characteristic of drift: the particle orbit is along the 
drift direction in regions of low field, and retrograde in regions of 
high field. The reentry points are therefore closer to the SGP than the 
exit points. Also note that there are almost no
 dots in this panel, i.e. reentry of particles with energy above 10$^{19.3}$ eV.
 Only lower energy particles scatter back inside 
 the 20 Mpc sphere. The top right-hand panel is for all particles
 that cross the 20 Mpc sphere in either direction.
 This is not the exact sum of the two left-hand panels since it includes 
particles exiting the sphere for the second (or third...) time. Also, 
since the statistical sample in each plot includes only the first 500 
occurencies for each panel, this distribution is not the sum of those
shown in the left two panels.

 The bottom righ-hand plot is for the protons that leave the 40 Mpc
 sphere and are not followed any more in the propagation Monte Carlo.
 Qualitatively this map is similar to the one above it, although
 the number of lower energy points is relatively smaller, since the
 lowest energy protons are dropped from the simulation as their total
 pathlength exceeds 400 Mpc. 
The bands of drifting particles have smaller $\phi$ offsets than for the 
top left panel since this angle is characteristically the width of the SGP 
divided by the radius of the simulation.

\begin{figure}[t] 
\centerline{\includegraphics[width=8.5cm]{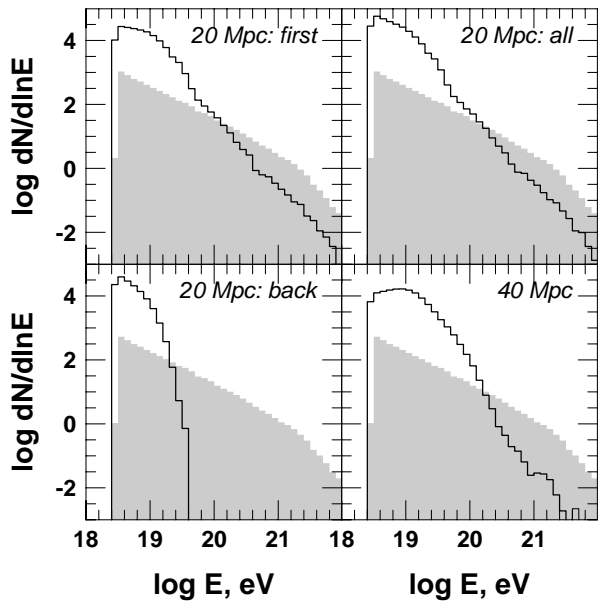}}
\caption{ Energy spectra of protons crossing the {\em front} cap
 in the same order as in Fig.~\protect\ref{exit_vec}.
 The shaded histogram gives the injection spectrum in the same
 solid angle. 
\label{ts63_spectra}
}
\end{figure}

  Fig.~\ref{ts63_spectra} shows the energy spectra of the particles that
 leave the spheres through the {\em front} cap under the same conditions.
 The top left-hand panel is for the particles that leave the 20 Mpc
 sphere for the first time - it has to be identical to the top left-hand
 panel in Fig.~\ref{fig3}, and it is. The bottom left-hand panel is for
 protons that backscatter into the 20 Mpc sphere through the {\em front} cap.
 There is a total cutoff of the energy spectrum at 10$^{19.6}$ eV.
 Most of the backscattered protons are of energy below 10$^{19}$ eV.
 The top right-hand panel shows the energy distribution of all protons
 crossing the {\em front} panel, independently of their direction.
 Since that distribution includes backscattered particles,
 the low energy flux is more enhanced than in the top left-hand
 panel here or in Fig.~\ref{fig3}.
 Finally the bottom right-hand panel gives the energy
 spectrum of the protons that leave the 40 Mpc sphere through its
 {\em front} cap. Although the spectrum has qualitatively the same
 features as for the 20 Mpc sphere, there are some differences.
 The first one is that the particle flux above 10$^{20}$ eV is here
 significantly lower. This is obviously a result of the longer propagation
 distance and correspondingly increased energy loss. There is also
 some redistribution in the energy range around 10$^{19}$ eV. Some of 
 the same particles that have populated the energy range above 10$^{20}$ eV
 have moved to this range, while some of the lower energy particles
 are lost from the simulation because of their large pathlength. 
 As a result, the energy spectrum is almost flat ($\sim E^{-1}$)
 in the range of 10$^{18.5}$-10$^{19.2}$

 Our conclusion derived from the results shown in
 Figs.~\ref{exit_vec}\&\ref{ts63_spectra} is that the account for
 the backscattered particles does not change qualitatively the
 energy spectra of the protons leaving the 20 Mpc sphere.
 Doubling the propagation distance to 40 Mpc creates changes
 in the energy spectra that are consistent with the increased
 proton energy loss. Qualitatively the energy spectra measured
 at the two distances show similar features depending on the
 exit position.

\subsection{External sources}

 A second concern is that the simulation does not account for possible 
sources outside the simulation volume. This is a complex problem 
involving the density of sources, their history and spectrum, etc. At 
high energy, one may presume that UHECRs will travel along nearly 
straight trajectories. For a mean separation of sources short
 compared to cosmological distances, one expects a homogeneous
 and isotropic phase space distribution of UHECRs.
 For lower energies, however, the particle 
 horizon may be limited by diffusion~\cite{Deligny},
 drift, or convection. In a matter 
dominated era, the distance between sources separates as $t^{2/3}$, where 
as the diffusive particle horizon grows only as $t^{1/2}$. It follows 
that we may observe sources in UHECRs but have no direct knowledge of the 
source spectra at energies below about $10^{19}$ eV. In fact, just such 
effects are seen in our main simulation. The front and back patches are 
on the same field line as the source, and see the full spectrum - even 
enhanced at low energies. Meanwhile the side and top patches are 
relatively devoid of low energy particles.

\begin{figure}[t] 
\centerline{\includegraphics[width=8.5cm]{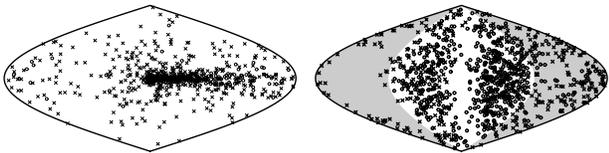}}
\caption{ Exit directions (left-hand panel) and
 positions (right-hand panel) of protons injected
 by an external source. The shaded area shows the hemisphere
 in which the protons enter the 20 Mpc sphere.
\label{fig:ts8pos}
}
\end{figure}

To mock up a distant external source, we expose our simulation volume to 
a broad beam of protons incident on the sphere from the negative 
$\hat{x}$ direction, i.e. the initial phase space distribution of the 
protons is $\hat{n} = (1,0,0)$ and evenly distributed across the disk of 
radius 20 Mpc, centered at $r_d = (-20,0,0)$~Mpc and normal to $\hat{n}$. 
As in the previous tests, the injected spectrum is flat
$E^{-1.25}$ so as to sample a 
wide variety of effects. We set $B=0$ in the region outside the 
simulation sphere ($r>20$~Mpc), so trajectories remain straight until 
they enter the simulation volume. 

Figure \ref{fig:ts8pos} shows the exit directions and positions. As 
usual, the high energy particles map out the geometry. The injection 
velocity $\hat{n}$ is at the center of the projection in panel a). 
Particles of high energy pass through the sphere with a modest amount of 
bending in the magnetic field, creating a tail which extends in the 
$-\phi$ direction. In panel b), the $+x$ hemisphere is in the center of 
the figure, with the center line locating the SGP on the exit side of the 
sphere. The SGP on the entry side corresponds to the outer boundary of 
the projection. High energy particles are seen to exit all on the side 
opposite from where they enter. Those which enter in the SGP are, for the 
most part, deflected into the $-y$ hemisphere. 

Lower energy particles show a different pattern of exit points. Those 
injected in the $+y$ hemisphere drift across the sphere and exit on the 
$+x$ side. Those injected with $y<0$ are in the region where drift is in 
the $-\hat{x}$ direction and, indeed, we see almost all low energy 
particles that exit with $y<0$ also exit with $x<0$. This is a 
fundamentally different behavior than is seen with the central source. In 
the central source simulation, particles were injected in the high field, 
but zero gradient, region at the center of the SGP. Low energy particles 
diffuse along the field lines, eventually exiting from either the front 
or back of the simulation volume. Only a few particles have initial 
energies and trajectories such that their motion is dominated by drift. 
For the  external source, most particles are injected into the moderate 
field, large gradient, regions outside the SGP, where low energy particle 
motion is dominated by drift, as opposed to diffusion. A few low energy 
particles injected into the SGP, i.e. $|y|<1.5$ Mpc, tend to diffuse 
along field lines and also exit within the SGP, explaining the halo of 
exit points at the edge of the projection.

\begin{figure}[t] 
\centerline{\includegraphics[width=8.5cm]{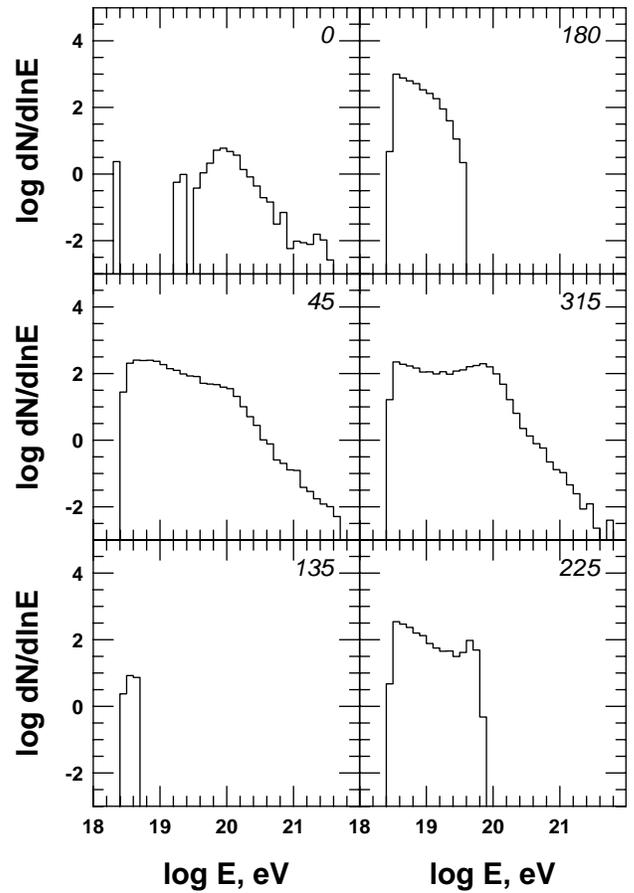}}
\caption{ Energy spectra of protons crossing the patches 
 described in the text, which are indicated in the panels
 with the value of the angle \protect$\phi$.
\label{fig:ts8spec}
}
\end{figure}

Figure \ref{fig:ts8pos} shows, once again, that in the presence of 
coherent magnetic fields, the flux density and spectrum observed depend 
on the location of the observer. To illustrate this further, in Figure 
\ref{fig:ts8spec} the spectrum of particles exiting the sphere through 
six observer patches is shown. Since a 
comparison of Figs.~\ref{fig:ts8pos} and \ref{caps_intro}
 indicates that our 
 previous set of patches are not located in regions of high particle flux, 
 we define six patches in the $z=0$ plane which we will call by the
 value of the $\phi$ angle that they correspond to.
 Patch {\em 0} is
 centered at $(x,y) = (20,0)$ and the spectrum consists only
 of UHE protons that manage to penetrate all the way through the SGP.
 Patch {\em 180} $(x,y) = (-20,0)$ accepts mainly low energy
 particles injected into the SGP which exit after 1/2 gyroorbit
 in the SGP field.
 Patch {\em 45} $(x,y) = (20/\sqrt{2},20/\sqrt{2})$ accepts both
 low energy particles that drift forward across the simulation
 volume and high energy particles that suffer mild deflections.
 The spectrum is similar to the injection spectrum.
 Patch {\em 225} $(x,y) = (-20/\sqrt{2},-20/\sqrt{2})$ accepts
 mainly low  energy particles which after injection drift back
 to exit points near their entry point.
 Patch {\em 315} $(x,y) = (20/\sqrt{2},-20/\sqrt{2})$ shows  an excess of
 high energy particles since high energy particles around 10$^{20}$ eV
 swept out of the SGP end up in this quadrant.
 Patch {\em 135}  $(x,y) = (-20/\sqrt{2},20/\sqrt{2})$ accepts neither
 high energy or low energy particles and the spectrum is suppressed at
 all energies, for particles exiting the simulation volume.

None of the spectra are particularly unusual, once the field geometry is 
accounted for, but they reinforce the conclusion that different 
observers will measure different spectra depending on their observation 
point relative to the local field geometry. The details depend on 
the source, as is seen by the different emphasis placed on diffusion and 
drift in the two simulations, but generically one requires a knowledge of 
field configuration in order to infer properties of the source from local 
observations of the particle flux.

\subsection{Impulse vs.\ steady state}

Our model assumes a steady state source, however, many models for UHECR 
sources are variable, episodic, or one shot explosions. It seems clear 
that any energy dependent transport process will result in an 
instantaneous observed spectrum that differs from the source spectrum. To 
study this we return to the central source simulation and examine the 
time delays as particles exit the simulation. The time delay is defined 
as $t_{d} = t_{exit} - t_0 - 20 {\rm Mpc/c}$, with $t_0$ being the
injection time.
\begin{figure}[t] 
\centerline{\includegraphics[width=8.5cm]{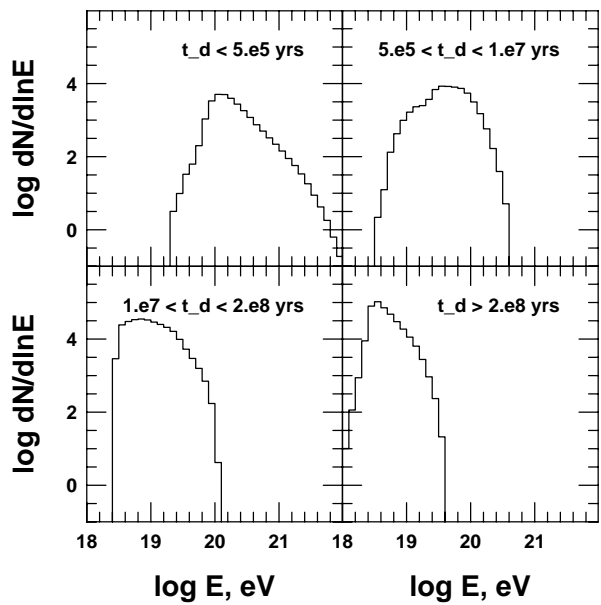}}
\caption{ Energy spectra of protons exiting the 20 Mpc sphere with
 different time delay as indicated in the panels.
\label{fig:ts3tdelay}
}
\end{figure}

 Figure~\ref{fig:ts3tdelay} shows the spectra of particles that exit the 
20 Mpc sphere in four bins of $t_d$. Generally, particles that exit promptly 
are those with high energy, where the delays are small and only due to 
slight bending of the particle trajectories in the simulation's magnetic 
field. Particles that exit late are particles of low energy, which have 
enhanced path lengths within the simulation during their diffusive 
transport. Figure \ref{fig:ts3tdelay} shows the spectra averaged over the 
whole sphere. More detailed patterns can be observed patch by patch. The 
main point here is that different observers, here separated in time, will 
observe different spectra. As before, we conclude that a program to turn 
observational data into statements of source spectra must take into 
account the possibility of organized extragalactic magnetic fields.

\section{Discussion and conclusions}

 We propagate protons of energy above 10$^{18.5}$ eV in the presence
 of regular and random extragalactic magnetic fields.
 The propagation is limited to the cosmologically small distances
 of 20 to 40 Mpc. We chose to have a single particle source in order
 to achieve a better understanding of the proton propagation.
 The observers in this scheme are located on the surface of 
 a 20 Mpc or a 40 Mpc sphere with the source in the center.
 The regular magnetic field is configured along a 3 Mpc wide
 `supergalactic plane'.

 Our general conclusions are that the particle densities inside the
 20 Mpc sphere, as well as the particle fluxes leaving the 20 and 40
 Mpc  spheres,
 depend strongly on the relative positions of the observers to
 the magnetic field directions. Lower energy particles are captured
 and channeled through the `supergalactic plane' both in the direction
 parallel and antiparallel to the magnetic field.

 To quantify better the energy spectra of the protons leaving the 20
 Mpc sphere in different locations we position several caps at
 different angles and distances from the magnetic wall. At the exit 
 from the 20 Mpc sphere inside the magnetic wall and on
 the same field line as the source (the caps {\em front} and {\em back} in
 Fig.~\ref{fig3}) the proton flux at around 10$^{19}$ eV exceeds
 the injection spectrum by almost two orders of magnitude.

 For particles exiting through caps that
 are at positions normal to
 the magnetic field direction and the SGP ({\em side} and {\em yz}), 
there is a lack of lower energy
 particles. Those of energy below about 10$^{19.3}$ eV, that
 are enhanced in the first two caps, are depleted since they lack the 
 magnetic rigidity to cross field lines to reach these caps.   

 The energy spectra in caps {\em top} and {\em xz} show intermediate 
spectra. Although direct propagation into these patches is strongly 
suppressed at low and intermediate energy, these caps capture a 
population of particles that drift along the edges of the SGP. Overall
we could not find a position on the 20 Mpc sphere where the energy
 spectrum of the exiting particles resembled the injection spectrum.

 The fluxes of particles above 10$^{20}$ eV are only mildly affected 
 by the position of the observer. Only a small fraction of these particles
 are contained in the `supergalactic plane' and the corresponding flux
 enhancement is minimal. Although reproducing the source beam pattern, 
the spectrum in this region is strongly affected by energy loss 
processes 
due to scattering off the cosmic microwave background. 

 The change in the flux density does not translate into anisotropic
 angular distribution for the particles that leave the 20 Mpc sphere
 along the magnetic field lines. The lower energy particles that are
 contained by the magnetic structure gyrate around the field lines
 and arrive at the {\em front} cap with almost isotropic angular
 distribution. Some degree of anisotropy can be seen at other locations
 that are reached only by higher energy protons. Even there the
 particle arrival angles do not coincide with the source direction
 and are strongly influenced by secondary propagation effects,
 such as drifts.

 In order to study the possible `edge effects' related to the short
 propagation distance and the existence of only a single source,
 we propagate protons in a larger, 40 Mpc sphere, concentric to
 the primary one. All protons that cross the 20 Mpc sphere,
 independent of their direction, are recorded, as well as
 those that leave the 40 Mpc sphere. The comparison between the
 positions in which the protons leave the two spheres and the
 energy spectra at the sphere crossings show qualitatively
 similar pictures. In both cases the lower energy particles 
 are confined to the `supergalactic plane' where the particle fluxes
 are significantly higher. The addition of the backscattered protons
 to the {\em front} cap at 20 Mpc increases the 10$^{19}$ eV excess
 over the injection spectrum. The energy spectrum of protons that
 leave the 40 Mpc hemisphere through its {\em front} cap is 
 influenced by the additional propagation distance, but with similar
 qualitative features.

 We also study the effects of the magnetic field on protons injected 
 by an external source by simulating the propagation of a plane front 
 of protons moving in direction $\hat{x}$ = (1,0,0). The quantitative
 effects in this scenario are different from these of a central source,
 but they are still caused by the proton motion in the magnetic field
 and generate significant angular deflection and changes in the
 `detected' proton energy spectra. 

 If the UHE protons are injected by an impulsive process, such
 as a gamma ray burst in the center of the 20 Mpc sphere, the
 proton exit spectrum depends heavily on the time delay after
 the burst. The fast particles arrive first, while the low energy
 ones suffer delays up to, and occasionally exceeding 1.3$\times$10$^{9}$
 yrs.   

 There are three reasons for performing the research that 
 we described. First, the presence or absence of a GZK cut-off
 is considered fundamental to understanding the sources and
 propagation of UHECR. The results presented here suggest that
 the possible presence of large scale 10 nG fields complicates
 the interpretation of UHECR data.
 The second one is related to the derivation of 
 the UHECR source luminosity, which is usually done in the
 vicinity of 10$^{19}$ eV. The simulations that we performed 
 emphasize the importance of accounting for the possible existence
 of ordered extragalactic magnetic fields. In their presence, the
 energy spectrum of particles emitted from a cosmologically nearby
 source depends very strongly on the relative position of the
 source and 
 observer to the direction of the ordered magnetic field.
 Neglecting these fields could lead to big errors in the estimate
 of the UHECR luminosity.

 The third reason is a study of the correlation between particle
 fluxes and arrival direction distribution. Even in the case of a single
 UHECR source, which we discuss in this paper, we could not find such
 a correlation, except for the highest energy protons.
 The enhanced particle fluxes in directions parallel
 to the magnetic field are almost isotropic, because of the 
 gyration of the protons around the magnetic field lines. An observer
 in the {\em front} cap would not be able to recognize the direction
 of our single source, except in the case of very large experimental
 statistics. Only by the observations of cosmic rays of energy
 exceeding 10$^{20}$ eV one would be able to see some clustering around
 the source direction.

 The dividing energy between diffusive and almost rectilinear propagation
 in our calculations is of order 10$^{19.3}$ eV. There are some 
 experimental indications, based on a part of the world
 UHECR statistics~\cite{Stanevetal95} that there is an increase in the
 anisotropy of these particles at energies above 10$^{19.6}$ eV. If this
 or a similar observation is confirmed in the future, it would
 suggest the existence of ordered magnetic fields of order
 20 nG in our cosmological neighbourhood.

 Our main conclusion is that the energy spectra and angular
 distribution   of protons accelerated both at a nearby or
 at a distant source are strongly affected  by  modest regular
 extragalactic magnetic fields.
 This is valid independently of the model for UHECR production. 
 Thus, it is important to allow for the possibility of large scale 
 magnetic fields in  UHECR data analysis and source searches.  On the 
 other hand, if a source of UHECR in a relatively wide energy range is 
 identified in the nearby Universe, and high experimental statistics exist 
 as expected from the currently active and future 
 experiments~\cite{HiRes,Auger,EUSO,OWL}, one could study the strength and 
 geometry of extragalactic magnetic fields that are not accessible in any 
 other way, utilizing techniques suggested by this research.

{\bf Acknowledgements} 
 The authors acknowledge fruitful discussions with P.L. Biermann, P.P.
 Kronberg, W. Matthaeus and J.W. Bieber. This research is supported
 in part by NASA Grant NAG5-10919.
 RE\&TS are supported also by the US Department of Energy contract
 DE-FG02 91ER 40626.

\bibliographystyle{prsty}

\end{document}